# Covariant formulation of at finite speed propagating electric interaction of moving charges in Euclidean geometry

Balázs Vető, retired associate professor of Eötvös University, Budapest, Hungary

*bzs.veto@gmail.com*   *veto@metal.elte.hu*

**Abstract**

Maxwell's electrodynamics postulates the finite propagation speed of electromagnetic (EM) action and the notion of EM fields, but it only satisfies the requirement of the covariance in Minkowski metric (Lorentz invariance). Darwin's force law of moving charges, which originates from Maxwell's field theory complies the Lorentz invariance as well. Poincaré's principle stating that physical laws can be formulated with identical meaning on different geometries suggest, that the retarded EM interaction of moving charges might be covariant even in Euclidean geometry (Galilean invariance). Keeping the propagation speed finite, but breaking with Maxwell's field theory in this study an attempt is made to find a Galilean invariant force law. Through the altering of the Liénard-Wiechert potential (LWP) a new retarded potential of two moving charges, the Common Retarded Electric Potential (CREP) is introduced which depends on the velocities of both interacting charges. The sought after force law is determined by means of the second order approximation of CREP. The law obtained is the Galilean invariant Weber's force law, surprisingly. Its rediscovery from the second order approximation of a retarded electric potential confirms the significance of Weber's force law and proves it to be a retarded and approximative law. The fact that Weber's force law implies even the magnetic forces tells us that magnetic phenomena are a manifestation of the retarded electric interaction exclusively. The third order approximation of the CREP opens the possibility of EM waves, and the creation of a complete, Euclidean electrodynamics.

## 1.    Poincaré's principle

The EM interaction of moving charged particles is described by the Darwin's force law in today physics. The Darwin's law comes from Maxwell's field theory and special relativity (SR). It takes the finite propagation speed of EM action into account but is not covariant in inertial frames of reference (IFR) using Euclidean geometry. Darwin's law was derived by means of a second order approximation of the LW potentials and it shows Lorentz invariance. Rather than the forces exerted by moving charges their Lorentz transformed values are identical in different IFR, because three-forces are not invariant measures in SR.

The question is raised, is SR the only possible physical treatment to find a covariant formulation of EM interaction with finite speed of propagation? Answering this question one should refer to the book [1] of the French physicist, mathematician and philosopher, H. Poincaré (1854-1912) titled „*La science et l'hypothese*" appeared in 1901. Poincaré explained that the geometry ordered to the physical space is not a property of the space, but a product of the human mind and it is arbitrarily adapted to the space. Hereby, different geometries can be ordered to the physical space. Physical laws exist in any geometry, while the mathematical form of the laws depends on the geometry. The physical meaning of the laws formulated on different geometries, is expected to be identical, but their mathematical form, the physical quantities' definitions affected in the laws may alter. Poincaré's principle suggests the answer that a covariant EM interaction might be formulated not only in Minkowski, but even in Euclidean geometry.

Inspired from Poincaré's principle a force law of EM interaction of moving charges is sought in this study. The force law sought after is based on the finite propagation speed of EM action, is expected to range with the experience and to give a covariant description of the EM forces in the frame of Euclidean geometry. Performing it, the LWP will be amended for the case of EM interaction of two moving charges instead of one moving and the other charge at rest situation.

The amended LWP breaks down with Maxwell's field theory. The LW retarded potential of a moving charge depends on the distance between the source and the observer, their relative displacement vector, as well as the velocity of the source charge, but is independent from the velocity of the observer. The new retarded potential (CREP) is a function of the velocities of both interacting charges.

Similarly, as Darwin's law was derived through the second order approximation of LWP, the desired force law will be obtained by means of the second order approximation of the CREP. In Poincaré's spirit, a force law equivalent with that of Darwin's one (2.6) is expected.

## 2. Retarded potentials and retarded forces

### 2.1 The Liénard-Wiechert potential (LWP) of a moving charge

A basic element of Maxwell's field theory is the taking into account of the finite propagation speed of the EM action. The finite propagation speed causes retardation in EM interactions. The way to the quantitative construction of retardation leads through the EM potentials. The retarded potential of an in the ether moving point charge with an observer at rest using Euclidean geometry was formulated by A. Liénard in 1898 [2] and E. Wiechert in 1900 [3], thus referred as Liénard-Wiechert potential.

Let $q_2$ be the point charge which generates the potential in the observation point $P_1$ and moving along the trajectory $\mathbf{r}_2(t)$. The position vector $\mathbf{r}_2(t)$ points from point $O$ to $q_2$. The constant position vector $\mathbf{r}_1$ points from point $O$ to $P_1$. See figure 1. The position vectors $\mathbf{r}(t) = \mathbf{r}_1 - \mathbf{r}_2(t)$ and $\mathbf{r}(t_r) = \mathbf{r}_1 - \mathbf{r}_2(t_r)$ point from $q_2$ to $P_1$ at $t$ and $t_r$, respectively. The action starts from $q_2$ at $t_r$, and reaches to $P_1$ just at the moment $t$.

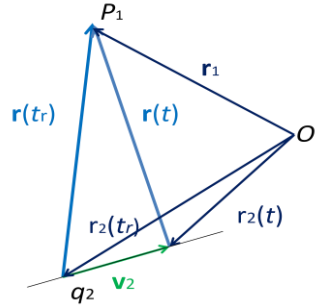

*Figure 1. Retarded LW potentials generated by $q_2$ in point $P_1$.*

The EM action needs time to reach from the source to the observation point. Using that the EM action propagates at the speed of light and considering that the motion of $q_2$ does not influence the light's propagation speed, the length of retardation time;

$$t - t_r = \frac{r(t_r)}{c}. \tag{2.1}$$

Because $\mathbf{r}_1$ is constant $d\mathbf{r}/dt = -\mathbf{v}_2$, and the LWP-s are in the point $P_1$ at the moment $t$:

$$\varphi_{2LW}(\mathbf{r}_1, t) = \frac{kq_2}{r(t_r) - \frac{\mathbf{v}_2(t_r)\mathbf{r}(t_r)}{c}}, \qquad \mathbf{A}_{2LW}(\mathbf{r}_1, t) = \varphi_{2LW}(\mathbf{r}_1, t)\frac{\mathbf{v}_2}{c^2} = \frac{kq_2}{r(t_r) - \frac{\mathbf{v}_2(t_r)\mathbf{r}(t_r)}{c}}\frac{\mathbf{v}_2}{c^2}. \tag{2.2}$$

In the denominator of LWP (2.2) there is a $-\mathbf{v}_2(t_r)\mathbf{r}(t_r)/c$ correction term for $r(t_r)$, for the distance between $q_2$ and $P_1$ at $t_r$. The correction term can be interpreted as the radial displacement of the source charge $q_2$ within the retardation time.

In order to have LWP only depend on the velocity of the source charge, one must set the observer at rest. Setting $\mathbf{v}_1 = 0$, keeps the retarded potential in the frame of Maxwell's field theory. As shown in [4], the correction term originates from the partial derivative;

$$\frac{\partial}{\partial t'}\left[t - \frac{1}{c}|\mathbf{r} - \mathbf{r}_s(t')|\right]_{t'=t_r} = \frac{1}{c}\frac{\mathbf{r} - \mathbf{r}_s(t')}{|\mathbf{r} - \mathbf{r}_s(t')|}\left(-\mathbf{v}_s(t')\right)_{t'=t_r}. \tag{2.3}$$

The vectors $\mathbf{r}$ and $\mathbf{r}_s(t')$ denote the positions of the observer and the source in [4]. When the motion of the observer is allowed then $\mathbf{r}$ becomes a function of $\mathbf{r}(t')$ and the derivative (2.3) changes into $[\mathbf{r}(t') - \mathbf{r}_s(t')][\mathbf{v}(t') - \mathbf{v}_s(t')]/[c|\mathbf{r}(t') - \mathbf{r}_s(t')|]|_{t'=t_r}$.

Back to our notations, the positions and velocities of the observer and the source are $\mathbf{r}_1; \mathbf{v}_1$, and $\mathbf{r}_2; \mathbf{v}_2$, respectively. Their relative positions and relative velocities are defined as $\mathbf{r}(t_r) = \mathbf{r}_1(t_r) - \mathbf{r}_2(t_r)$ and $\mathbf{v}(t_r) = \mathbf{v}_1(t_r) - \mathbf{v}_2(t_r)$. At last, the correction term is $\mathbf{v}(t_r)\mathbf{r}(t_r)/c$ in the case of a moving observer, instead of $-\mathbf{v}_2(t_r)\mathbf{r}(t_r)/c$ as given in (2.2).

When the observer moves, its velocity appears in the correction term of the retarded potential. A potential which involves the observers velocity is outside of Maxwell's field theory. Furthermore, in the case of a moving observer, the length of the retardation time given in (2.1) does also change.

After creating SR it turned out that the Lorentz-transformed Coulomb potential of a moving point charge leads to its LWP without any restriction on the observer's velocity. Therefore LWP became a way to SR connecting retardation and Lorentz transformation. LWP cannot be used in a Euclidean environment but it capable of describing the EM interaction of two moving charges in the frame of SR.

### 2.2 The second order approximation of LWP, an action-at-a-distance force from a retarded potential

The LWPs (2.2) are given in the observation point $P_1$ at the moment $t$, however are determined by the retarded measures on the RHS which belong to an earlier time $t_r$. If the quantities $\mathbf{r}(t_r)$, $r(t_r)$ and $\mathbf{v}_2(t_r)$ were known as functions of $\mathbf{r}(t)$, $r(t)$ and $\mathbf{v}_2(t_r)$, the $\varphi_{2LW}(\mathbf{r}_1, t)$ potential could be expressed by the measures valid at the moment $t$. This way $\varphi_{2LW}(\mathbf{r}_1, t)$ would formally become an action-at-a-distance potential. The measures $\mathbf{r}(t_r)$, $r(t_r)$ and $\mathbf{v}_2(t_r)$ cannot be expressed explicitly by $\mathbf{r}(t)$, $r(t)$ and $\mathbf{v}_2(t)$, instead they have to be developed into power series including the second order terms of $1/c$.

Omitting the argument of $(t)$ and writing simply $\mathbf{r}, r, \mathbf{v}_2$, and $v_2$ the second order approximation of the retarded scalar and vector potentials (2.2) is given in Landau and Lifschitz [5],

$$^2\varphi_{2LW} = \frac{kq_2}{r}\left[1 + \frac{v_2^2}{2c^2} - \frac{(\mathbf{r}\mathbf{v}_2)^2}{2c^2r^2}\right], \qquad ^2\mathbf{A}_{2LW} = \varphi_{2LW}\frac{\mathbf{v}_2}{c^2} \approx \frac{kq_2\mathbf{v}_2}{c^2r} = \mathbf{A}_2\,. \tag{2.4}$$

### 2.3  Retarded electromagnetic interaction of two point charges, Darwin's force law

The SR makes the LWP valid even for the case of a moving observer, i.e. $\mathbf{v}_1 \neq 0$ . C. G. Darwin investigated the EM interaction of moving charges in 1920 [6]. He proceeded from the second order approximation of LWP (2.4) to determine the retarded forces. Using SR and the second order approximation of LWP, Darwin formulated a Lagrangian of two moving charges. By adding the total time derivative of a function of the coordinates he got the Darwin's Lagrangian. The Lagrangian of a point mass $m_1$ carrying a charge $q_1$ moving at a speed of $\mathbf{v}_1$ in the EM field of at a speed $\mathbf{v}_2$ moving charge $q_2$;

$$^2L_{1D} = m_1\frac{v_1^2}{2} + m_1\frac{v_1^4}{8c^2} + \frac{kq_1q_2}{2rc^2}\left[\mathbf{v}_1\mathbf{v}_2 + \frac{(\mathbf{r}\mathbf{v}_1)(\mathbf{r}\mathbf{v}_2)}{r^2}\right] - \frac{kq_1q_2}{r}. \tag{2.5}$$

Darwin's force comes from the Euler-Lagrange equation as a second order, action-at-a-distance approximation of $^2\mathbf{F}_{1D} = d\,{}^2\mathbf{p}_1/dt$, where $^2\mathbf{p}_1 = m_1\left(1 + \frac{v_1^2}{2c^2}\right)\mathbf{v}_1$. The Darwin force exerted by $q_2$ on $q_1$;

$$^2\mathbf{F}_{1D} = \frac{kq_1q_2}{r^3}\mathbf{r}\left[1 + \frac{v_2^2}{2c^2} - \frac{3}{c^2r^2}\frac{(\mathbf{r}\mathbf{v}_2)^2}{c^2} - \frac{\mathbf{r}\mathbf{a}_2}{2c^2}\right] + \frac{kq_1q_2}{c^2r^3}\left[\mathbf{v}_1 \times (\mathbf{v}_2 \times \mathbf{r}) - \frac{\mathbf{a}_2 r^2}{2}\right]. \qquad (2.6)$$

The force law (2.6) is not covariant in Euclidean geometry. When comparing the forces $^2\mathbf{F}_{1D}$ and $^2\mathbf{F}_{2D}$ the law of action and reaction is not fulfilled. The magnetic force terms in $^2\mathbf{F}_{1D}$ and $^2\mathbf{F}_{2D}$ are not even parallel. The Lorentz transformed Darwin forces are covariant in Minkowski metric. Instead of the two forces, their Lorentz transformed values are equal in different inertial frames of reference, because three-forces are not an invariant measures in SR.

## 3.   Revision of LWP, a new approach to the retarded electric force between moving charges

### 3.1 The common retarded electric potential (CREP)

Inspired by Poincaré's principle a force law capable of describing the EM interaction of moving charges is sought. A force law that comes from a retarded potential, i.e. considers the finite propagation speed of electric action and is covariant in Euclidean geometry. As in section 2.1 was established, LWP is not able to provide any force law satisfying the above expectations. Therefore a new potential is needed. The new potential ordered to both of the interacting charges, will be formulated by the revision of LWP.

The formulation the CREP of the moving charges $q_1$ and $q_2$ is based on the LW scalar potential. The $\varphi_{2LW}(\mathbf{r}_1, t)$ given in (2.2) is generated by the moving charge $q_2$ and perceived by the charge $q_1$ which is at rest. In section 2.1 was shown, that in the case of a moving observer the observer's velocity appears in the retarded potential. The length of the retardation time changes as well.

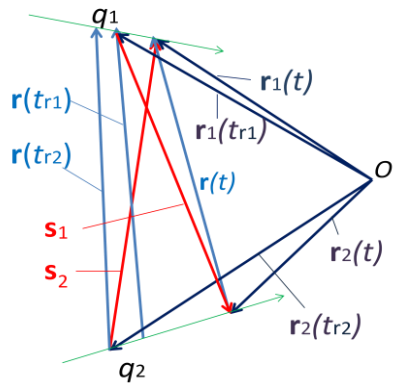

***Figure 2.** Geometric representation of two moving charges.*

The CRE potentials $\varphi_{2CR}[\mathbf{r}_1(t)]$ and $\varphi_{1CR}[\mathbf{r}_2(t)]$ which are generated and perceived by $q_1$ and $q_2$ at each-other location conversely will be formulated. The action perceived by the charges $q_1$ and $q_2$ at the moment $t$, starts from the sources at different $t_{r1}$ and $t_{r2}$ instants. The electric action traces different $s_1$ and $s_2$ distances coming along from the sources, $q_1$, $q_2$ to the perceivers $q_2$ and $q_1$, see figure 2. For the sake of symmetry the formulation of the potentials $\varphi_{2CR}[\mathbf{r}_1(t)]$ and $\varphi_{1CR}[\mathbf{r}_2(t)]$ follows the same steps, so it is enough to derive only one of them. Let us choose the CREP $\varphi_{2CR}[\mathbf{r}_1(t)]$ now.

To turn the LWP $\varphi_{2LW}(\mathbf{r}_1, t)$ into the CREP $\varphi_{2CR}[\mathbf{r}_1(t)]$ three amendments have to be carried out:

a) *Replacing the velocity of the generating charge with the relative velocity of the interacting charges.* When the perceiving charge $q_1$ is not required to be at rest, i.e. $\mathbf{v}_1 \neq 0$ then the relative velocity $\mathbf{v}(t_{r2}) = \mathbf{v}_1(t_{r2}) - \mathbf{v}_2(t_{r2})$ takes the place of $-\mathbf{v}_2(t_r)$ as shown in section 2.1. This modification is a simply correction of the arbitrary restriction in LWP, done in (2.3).

b) *Amendment the length of the retardation time.* The change of the length of the retardation time is a consequence of the motion of the observer, $q_1$. In the case of a moving $q_1$ the electric action performs the path $s_2(t_{r2})$ when propagating from $q_2$ to $q_1$. Here $\mathbf{s}_2(t_{r2}) = \mathbf{r}_1(t) - \mathbf{r}_2(t_{r2})$, see figure 2. Therefore $t - t_{r2} = s_2(t_{r2})/c$ takes the place of $t - t_r = r(t_r)/c$ in (2.2). This way we got a new retarded potential:

$$\varphi_{2NR}[\mathbf{r}_1(t)] = \frac{kq_2}{r(t_{r2}) + \frac{s_2(t_{r2})}{c} \frac{\mathbf{v}(t_{r2})\mathbf{r}(t_{r2})}{r(t_{r2})}}, \qquad \text{where } t_{r2} = t - \frac{s_2(t_{r2})}{c}. \tag{3.1}$$

c) *The averaging of the scalar product of* $\mathbf{v}(t_{r2})\mathbf{r}(t_{r2})$. The third modification is a refinement of the potential (3.1). The meaning of the second term in the denominator of (3.1) is the relative radial displacement of the charges during the time of $s_2(t_{r2})/c$. The product $\mathbf{v}(t_{r2})\mathbf{r}(t_{r2})$ is considered to be constant during the time of $s_2(t_{r2})/c$. The product indeed changes during the time. Therefore, to obtain a more accurate result, the scalar product $\mathbf{v}(t_{r2})\mathbf{r}(t_{r2})$ will be replaced by the arithmetic mean of the start and end values of the factors; $\bar{\mathbf{v}}[t_{r2}] = [\mathbf{v}(t_{r2}) + \mathbf{v}(t)]/2$ and $\bar{\mathbf{r}}[t_{r2}] = [\mathbf{r}(t_{r2}) + \mathbf{r}(t)]/2$.

Carrying out the above replacement in (3.1) both of the sought CREPs are obtained;

$$\varphi_{1CR}[\mathbf{r}_2(t)] = \frac{kq_1}{r(t_{r1}) + \frac{s_1(t_{r1})}{r(t_{r1})c} \bar{\mathbf{v}}[t_{r1}]\, \bar{\mathbf{r}}[t_{r1}]}; \quad \varphi_{2CR}[\mathbf{r}_1(t)] = \frac{kq_2}{r(t_{r2}) + \frac{s_2(t_{r2})}{r(t_{r2})c} \bar{\mathbf{v}}[t_{r2}]\, \bar{\mathbf{r}}[t_{r2}]}. \tag{3.2}$$

The CRE potentials $\varphi_{1CR}[\mathbf{r}_2(t)]$ and $\varphi_{2CR}[\mathbf{r}_1(t)]$ formulated in (3.2) are retarded electric potentials of two moving charges. We expect these potentials to provide a correct and covariant description of the electric interaction of moving charges in Euclidean geometry. The potentials (3.2) involve the relative velocities of the two interacting charges as well as the absolute velocity of the generating charges, because $s_1(t_{r1})$ and $s_2(t_{r2})$ depend each on the $\mathbf{v}_1(t_{r1})$ and $\mathbf{v}_2(t_{r2})$ velocities of the source charge, respectively. Therefore CREP (3.2) is not a Galilean invariant potential. This attribute of the CREP should not discourage us from the investigation of its second order approximation.

*3.2 Second order, action-at-a-distance approximation of the CRE potential of two moving charges*

The CRE potential $\varphi_{2CR}[\mathbf{r}_1(t)]$ involves the variable, $t_{r2}$ on the RHS of equation (3.2). The retarded quantities of $\bar{\mathbf{r}}(t_{r2})$, $r(t_{r2})$, $\bar{\mathbf{v}}(t_{r2})$ and $s_2(t_{r2})$ have to be expressed by means of their values at the moment of perception $\mathbf{r}(t)$, $\mathbf{v}(t)$, $r(t)$, $\mathbf{v}_2(t)$, as was done in the case of LWP in section 2.2. This transformation of variables will be performed by developing the $\bar{\mathbf{r}}(t_{r2})$, $r(t_{r2})$, $\bar{\mathbf{v}}(t_{r2})$ and $s_2(t_{r2})$ quantities into power series with the second order term of $1/c$ inclusive. The derivation of the second order approximation of CREP is located in the Appendix.

Omitting the argument $(t)$ in the $\mathbf{r}(t)$, $\mathbf{v}(t)$, $r(t)$ functions the second order approximation of the CREPs;

$$^2\varphi_{1CR}[\mathbf{r}_2(t)] = \frac{kq_1}{r}\left[1 - \frac{(\mathbf{r}\mathbf{v})^2}{2r^2c^2}\right], \quad \text{and} \quad ^2\varphi_{2CR}[\mathbf{r}_1(t)] = \frac{kq_2}{r}\left[1 - \frac{(\mathbf{r}\mathbf{v})^2}{2r^2c^2}\right], \tag{3.3}$$

as obtained in equation (A.12). The terms involving the absolute velocities of $\mathbf{v}_1$ and $\mathbf{v}_2$ are canceled out. Only the relative positions and velocities remain in the second order approximation. Therefore the second order potentials (3.3) are Galilean invariant.

Now, it is possible to write the retarded electric interaction energy of the moving charges, as

$$U(\mathbf{r}, \mathbf{v}) = q_1 \, {}^2\varphi_{2R}[\mathbf{r}_1(t)] = q_2 \, {}^2\varphi_{1R}[\mathbf{r}_2(t)] = \frac{kq_1q_2}{r}\left[1 - \frac{(\mathbf{r}\mathbf{v})^2}{2r^2c^2}\right]. \tag{3.4}$$

(3.4) is identical with the Weber's energy, see Assis and Torres [7]. The derivative of the interaction energy with respect to the position vector of the object on which the force is acting provides the force law,

$$^2\mathbf{F}_{1CR} = -\frac{dU(\mathbf{r}, \mathbf{v})}{d\mathbf{r}_1}, \qquad {}^2\mathbf{F}_{2CR} = -\frac{dU(\mathbf{r}, \mathbf{v})}{d\mathbf{r}_2}. \tag{3.5}$$

The force law obtained is a second order, action-at-a-distance approximation. Because, $\mathbf{r} = \mathbf{r}_1 - \mathbf{r}_2$, it is obvious, that $\frac{d}{d\mathbf{r}_1} = \frac{d}{d\mathbf{r}}$, and $\frac{d}{d\mathbf{r}_2} = -\frac{d}{d\mathbf{r}}$. The derivation $\frac{dU(\mathbf{r}, \mathbf{v})}{d\mathbf{r}}$ is easy to perform as, $\frac{dU(\mathbf{r}, \mathbf{v})}{d\mathbf{r}} = \frac{dr}{d\mathbf{r}}\frac{dt}{dr}\frac{dU(\mathbf{r}, \mathbf{v})}{dt}$. The sought force law obtained is the well known Weber's force law;

$$^2\mathbf{F}_{1CR} = -\frac{dU(\mathbf{r}, \mathbf{v})}{d\mathbf{r}_1} = \frac{kq_1q_2}{r^3}\mathbf{r}\left[1 + \frac{v^2}{c^2} - \frac{3(\mathbf{v}\mathbf{r})^2}{2r^2c^2} + \frac{\mathbf{a}\mathbf{r}}{c^2}\right] \equiv \mathbf{F}_{1W}. \tag{3.6}$$

The same result comes out by means of the Euler-Lagrange equation, when considering $U = kq_1q_2/r$ to be the potential and $T = -kq_1q_2(\mathbf{r}\mathbf{v})^2/(2r^3c^2)$ the kinetic part of the (3.4) interaction energy and the Lagrangian is given by $L = T - U$.

The sought after force law of the EM interaction of two moving charges is derived from a retarded, electric potential, CREP in (3.6), which takes the finite propagation speed of EM action into account and is covariant in Euclidean geometry. Surprisingly, the force law obtained is the Weber's force law. W. Weber built his law upon experimental data in 1846 [8]. Above derivation gives Weber's force law a convincing theoretical background, a solid claim by implying the finite propagation speed of EM action.

Weber's force law disappeared from the textbooks and from the mind of physicists' society after the 1920's. After the discovery of EM waves the idea of action-at-a-distance became untenable further Darwin's and Weber's force laws provide different results of EM interaction of moving charges. A wide overview on the Weber's force law and its applications is given by Assis [9].

## 4. Conclusions

### 4.1 Weber's force law takes the finite propagation speed of electric interaction into account

A Galilean invariant force law of electrically interacting moving charges was derived in this study. The obtained law is the Weber's force law. Weber originally considered his law to describe an action-at-a-distance EM interaction. Our derivation proves that Weber's force law originates from the second order approximation of a retarded potential, CREP, like Darwin force originates from LWP. Although Weber considered his law to be an exact relationship it is an approximation of order $1/c^2$. The theoretical rediscovery of Weber's force law is a remarkable epistemological result. Present study raised Weber's great force law to its rightful place with a delay of 175 years.

### 4.2 The magnetism is a simply consequence of the finite propagation speed of the electric action

An only electric, retarded force of moving charges is derived in this study.  Weber's force law however, involves the magnetic forces as well. There is no need to introduce extra magnetic vector potential to find magnetic forces. Magnetic forces come simply from the retarded electric potential. Similarly in SR, the EM four-potential of a charge at rest is its Coulomb potential. When a charge is moving, its four-potential is given by the Lorentz transformation of the rest four-potential. The transformed four-potential involves the magnetic vector potential of the charge automatically.  We assert that magnetism is the manifestation of the retarded electricity with no respect of the used geometry.

*4.3 Weber's force law, the third order approximation of CREP and EM waves*

Weber's force law in its original conception seems to be an impasse because the action-at-a-distance contradicts the existence of EM waves. Some efforts were made by different authors to extend Weber's force law for radiation of EM waves. Wesley [10] for example introduced two additional potentials to the usual electric and magnetic potentials to solve the problem of EM radiation.

In present study, the existence of EM waves is self-explanatory. The second order approximation of CREP leads to a Galilean invariant potential and provides Weber's force law. The third order terms in a higher approximation of CREP implies Maxwell type potentials which depend on the velocity and acceleration of the source charge only. These Maxwell type third order terms maintain EM waves, as preliminary calculations indicate. The CREP introduced in this paper offers the possibility to develop a complete electrodynamics which implies the Galilean invariant EM interaction of moving charges and the existence of EM waves.

*4.4 Comparison of Darwin's and Weber's force and Poincaré's principle*

The Galilean invariant Weber's and Lorentz invariant Darwin's force laws provide different EM force between moving charges, see for example Assis [11], or by comparing the equations of (2.6) and (3.6). A Lorentz invariant and a Galilean invariant force can be compared in the rest frame of the object affected by the force only. In our case this object is the charge of $q_1$. The rest frame of $q_1$ is defined by the condition, $\mathbf{v}_1 = 0$. Replacing $\mathbf{v}_1 = 0$ and $\mathbf{a}_1 = 0$ into equations (3.6) and (2.6) the two forces take the form of;

$$\mathbf{F}_{1W} = \frac{kq_1q_2}{r^3}\mathbf{r}\left[1 + \frac{v_2^2}{c^2} - \frac{3}{2}\frac{(\mathbf{r}\mathbf{v}_2)^2}{c^2r^2} - \frac{\mathbf{a}_2\mathbf{r}}{c^2}\right]; \qquad \mathbf{F}_{1D} = \frac{kq_1q_2}{r^3}\left\{\mathbf{r}\left[1 + \frac{v_2^2}{2c^2} - \frac{3}{2}\frac{(\mathbf{r}\mathbf{v}_2)^2}{c^2r^2} - \frac{\mathbf{a}_2\mathbf{r}}{2c^2}\right] - \frac{\mathbf{a}_2r^2}{2c^2}\right\}. \quad (4.1)$$

The two forces are different in the rest frame of $q_1$. Both force laws cannot be correct. Only experiments can tell which force law is the right one.

Poincaré's principle is not confirmed in present study because Darwin's force law and the rediscovered Weber's force law are not equivalent. At last we conclude that there exist a covariant description of the EM interaction of moving charges implying the finite propagation speed, without using Maxwell's field theory and special relativity.

**Appendix –** *The second order approximation of CREP*

The retarded quantities $\bar{\mathbf{r}}(t_{r2})$, $r(t_{r2})$, $\bar{\mathbf{v}}(t_{r2})$ and $s_2(t_{r2})$ have to be expressed by means of their values at $t$, at the moment of perception, i.e. by $\mathbf{r}(t)$, $\mathbf{v}(t)$, $r(t)$, $\mathbf{v}_2(t)$. Here $t_{r2} = t - s_2(t_{r2})/c$ as it follows from (3.1). Developing the retarded quantities into power series at the time $t$ inclusively the second order terms of $1/c$; $\qquad f(t_{r2}) = f(t) + \dot{f}(t)(t_{r2} - t) + \ddot{f}(t)(t_{r2} - t)^2/2.$

Because of the symmetry of the subscripts it is sufficient to carry out the approximation for one of the CRE potentials given in equation (3.2). In this section the potential $\varphi_{2CR}[\mathbf{r}_1(t)]$ is chosen;

$$\varphi_{2CR}[\mathbf{r}_1(t)] = \frac{kq_2}{r(t_{r2}) + \frac{s_2(t_{r2})}{r(t_{r2})c}\bar{\mathbf{v}}[t_{r2}]\,\bar{\mathbf{r}}[t_{r2}]}. \qquad (A.1)$$

Let us first write the second order approximation of the denominator of (A.1):

$$^2N_{2R} = {}^2r(t_{r2}) + \frac{1}{c}{}^1\left(\frac{s_2(t_{r2})}{r(t_{r2})}\bar{\mathbf{v}}[t_{r2}]\,\bar{\mathbf{r}}[t_{r2}]\right), \qquad (A.2)$$

and calculate the intended approximation of all quantities with argument $t_{r2}$ in (A.2).

The second order approximation of $^2r(t_{r_2})$ and first order of the other quantities is needed. The basic relations follow from the geometry, $\mathbf{r}(t_{r_2}) = \mathbf{r}_1(t_{r_2}) - \mathbf{r}_2(t_{r_2})$; and $\mathbf{s}_2(t_{r_2}) = \mathbf{r}_1 - \mathbf{r}_2(t_{r_2})$. The argument $(t)$ is omitted. A simple consequence of basic relations, that $\mathbf{s}_2(t) = \,^0\mathbf{s}_2(t_{r_2}) = \mathbf{r}$ and $s_2(t) = \,^0s_2(t_{r_2}) = r$. The first order approximation of $\mathbf{r}(t_{r_2})$;

$$^1\mathbf{r}(t_{r_2}) = \mathbf{r}_1 - \mathbf{v}_1 \frac{^0s_2(t_{r_2})}{c} - \mathbf{r}_2 + \mathbf{v}_2 \frac{^0s_2(t_{r_2})}{c} = \mathbf{r} - \mathbf{v} \frac{^0s_2(t_{r_2})}{c} = \mathbf{r} - \mathbf{v}\frac{r}{c}. \tag{A.3}$$

Similarly;
$$^1\mathbf{v}(t_{r_2}) = \mathbf{v} - \mathbf{a}\frac{^0s_2(t_{r_2})}{c} = \mathbf{v} - \mathbf{a}\frac{r}{c} \tag{A.4}$$

$$^1\mathbf{s}_2(t_{r_2}) = \mathbf{r}_1 - \,^1\mathbf{r}_2(t_{r_2}) = \mathbf{r} - \mathbf{v}_2\,^0s_2(t_{r_2})/c = \mathbf{r} - \mathbf{v}_2\frac{r}{c}. \tag{A.5}$$

Hereafter the quantity $^1s_2(t_{r_2})$ is calculated, using $\dot{\mathbf{s}}_2(t) = -\mathbf{v}_2$ is given,

$$^1s_2(t_{r_2}) = s_2(t) - \dot{s_2}(t)\frac{^0s_2(t_{r_2})}{c} = r - \frac{\mathbf{s}_2(t)\dot{\mathbf{s}}_2(t)}{s_2(t)c}r = r + \frac{\mathbf{r}\mathbf{v}_2}{c} \tag{A.6}$$

Similarly will be expressed of the second order approximation of the distance $^2r(t_{r_2})$ by quantities depend on $t$. Developing the power series and replacing $^1s_2(t_{r_2})$ from (A.6) and $^0s_2(t_{r_2}) = r$;

$$^2r(t_{r_2}) = r - \frac{\dot{r}\,^1s_2}{c} + \frac{\ddot{r}\,^0s_2^2}{2c^2} = r\left[1 - \frac{\mathbf{r}\mathbf{v}}{rc} + \frac{v^2}{2c^2} - \frac{(\mathbf{r}\mathbf{v})(\mathbf{r}\mathbf{v}_2)}{r^2c^2} - \frac{(\mathbf{r}\mathbf{v})^2}{2r^2c^2} + \frac{\mathbf{r}\mathbf{a}}{2c^2}\right]. \tag{A.7}$$

From (A.7) follows $^1r(t_{r_2}) = r\left(1 - \frac{\mathbf{r}\mathbf{v}}{rc}\right)$, $^0r(t_{r_2}) = r$.

At last the first order approximation of the mean values $\bar{\mathbf{v}}[t_{r_2}]$; $\bar{\mathbf{r}}[t_{r_2}]$ are determined;

$$^1\bar{\mathbf{v}}[t_{r_2}] = \frac{^1\mathbf{v}(t_{r_2}) + \mathbf{v}(t)}{2} = \mathbf{v} - \mathbf{a}\frac{r}{2c}; \qquad ^1\bar{\mathbf{r}}[t_{r_2}] = \frac{^1\mathbf{r}(t_{r_2}) + \mathbf{r}(t)}{2} = \mathbf{r} - \mathbf{v}\frac{r}{2c} \tag{A.8}$$

Hereby all the quantities needed in the denominator (A.2) are determined. Inserting them into (A.2);

$$^2N_{2R} = r\left[1 - \frac{\mathbf{r}\mathbf{v}}{rc} + \frac{v^2}{2c^2} - \frac{(\mathbf{r}\mathbf{v})(\mathbf{r}\mathbf{v}_2)}{r^2c^2} - \frac{(\mathbf{r}\mathbf{v})^2}{2r^2c^2} + \frac{\mathbf{r}\mathbf{a}}{2c^2}\right] + \frac{1}{c}\,^1\!\left[\frac{\left(\mathbf{v} - \mathbf{a}\frac{r}{2c}\right)\left(\mathbf{r} - \mathbf{v}\frac{r}{2c}\right)r\left(1 + \frac{\mathbf{r}\mathbf{v}_2}{rc}\right)}{r\left(1 - \frac{\mathbf{r}\mathbf{v}}{rc}\right)}\right]. \tag{A.9}$$

Developing of the first order approximation of the second term of (A.9),

$$\frac{1}{c}\,^1\!\left[\left(\mathbf{v} - \mathbf{a}\frac{r}{2c}\right)\left(\mathbf{r} - \mathbf{v}\frac{r}{2c}\right)\left(1 + \frac{\mathbf{r}\mathbf{v}_2}{rc}\right)\left(1 + \frac{\mathbf{r}\mathbf{v}}{rc}\right)\right] = \frac{\mathbf{r}\mathbf{v}}{c} - r\frac{\mathbf{r}\mathbf{a}}{2c^2} - r\frac{v^2}{2c^2} + \frac{(\mathbf{r}\mathbf{v})^2}{rc^2} + \frac{(\mathbf{r}\mathbf{v})(\mathbf{r}\mathbf{v}_2)}{rc^2}. \tag{A.10}$$

Replacing (A.10) into (A.9), the second order approximation of the denominator (A.2) is obtained;

$$^2N_{2R} = r\left[1 - \frac{\mathbf{r}\mathbf{v}}{rc} + \frac{v^2}{2c^2} - \frac{(\mathbf{r}\mathbf{v})(\mathbf{r}\mathbf{v}_2)}{r^2c^2} - \frac{(\mathbf{r}\mathbf{v})^2}{2r^2c^2} + \frac{\mathbf{r}\mathbf{a}}{2c^2} + \frac{\mathbf{r}\mathbf{v}}{rc} - \frac{\mathbf{r}\mathbf{a}}{2c^2} - \frac{v^2}{2c^2} + \frac{(\mathbf{r}\mathbf{v})^2}{r^2c^2} + \frac{(\mathbf{r}\mathbf{v})(\mathbf{r}\mathbf{v}_2)}{r^2c^2}\right] = r\left(1 + \frac{(\mathbf{r}\mathbf{v})^2}{2r^2c^2}\right) \tag{A.11}$$

At last, the second order approximation of the CRE potential $^2\varphi_{2CR}[\mathbf{r}_1(t)]$ is given as:

$$^2\varphi_{2CR}[\mathbf{r}_1(t)] = \frac{kq_2}{^2N_{2R}} = \frac{kq_2}{r\left[1 + \frac{(\mathbf{r}\mathbf{v})^2}{2r^2c^2}\right]} = \frac{kq_2}{r}\left[1 - \frac{(\mathbf{r}\mathbf{v})^2}{2r^2c^2}\right]. \tag{A.12}$$

## Acknowledgements

I am grateful prof. Assis for calling my attention to Weber's electrodynamics and for the impressive conversation when met in Budapest.

## References

[1]   H. Poincaré, 1901. *Science and Hypothesis.* W. Scott, London, 1905. Original published in French, 1901.

[2]   A. Liénard, 1898. *Champ electrique et magnetique produit par une charge concentree en un point et animee d'un mouvement queoconque.* L'eclairage Electrique, vol. 16. pp 5-106

[3]   E. Wiechert, 1901. *Elektrodynamische Elementargesetze.* Annalen der Physik, vol. 309.pp. 667-689

[4] https://en.wikipedia.org/wiki/Li%C3%A9nard%E2%80%93Wiechert_potential

[5]   L.D. Landau and E.M. Lifschitz, 1975. *The Classical Theory of Fields.* Butterworth-Heinemann, Burlington. Fourth revised English Edition.

[6]   C.G. Darwin, 1920. *The dynamical motions of charged particles.* Philosophical Magazine, vol. 39. pp. 537-551

[7]   A.K.T. Assis and H. Torres Silva, 2000. *Comparison between Weber's electrodynamics and classical electrodynamics,* Indian Academy of Sciences, Pramana – Journal of Physics, vol. 55. No. 3. pp 393-404.

[8]   W. Weber, 1846. *Elektrodynamische Maassbestimmungen – Über ein allgemeines Grundgesetz der elektrischen Wirkung.* Abhandlungen bei Begründung der Königl. Sächs. Gesellschaft der Wissenschaften am Tage der zweihundertjärigen Geburtstagfeier Leibnizen's. Herausg. von der Fürstl. Jablonowskischen Gesellschaft, Leipzig. pp. 211-378.

[9]   A.K.T. Assis, 2014. *Relational Mechanics and Implementation of Mach's Principle with Weber's Gravitational force.* Aperion, Montreal. C. Roy Keys Inc.

[10]  J.P. Wesley, 1990. *Weber Electrodynamics,* Foundations of Physics Letters, Vol. 3. No. 5. pp. 443-469.

[11]  A.K.T. Assis, 1995. *Weber's force versus Lorentz's force*, Phys. Essays Vol. 8. pp 335-341.